\documentclass  [aps,prl,twocolumn,byrevtex,showpacs,preprintnumbers,amsmath,amssymb,superscriptaddress,floatfix,longbibliography]{revtex4-2}

\usepackage{graphicx}
\usepackage{color}
\usepackage{kotex}

\begin{document}
\title{Comprehensive peak-width analysis in matter-wave diffraction under grazing incidence conditions}

\author{Lee Yeong Kim}
\affiliation{Department of Physics, Ulsan
National Institute of Science and Technology (UNIST), Ulsan 44919, Republic of Korea}
\author{Do Won Kang}
\email{dowon5961@unist.ac.kr}
\affiliation{Department of Chemistry, Ulsan
National Institute of Science and Technology (UNIST), Ulsan 44919, Republic of Korea}
\author{Jong Chan Lee}
\affiliation{Department of Chemistry, Ulsan National Institute of Science and Technology (UNIST), Ulsan 44919, Republic of Korea}
\author{Eunmi Chae}
\affiliation{Department of Physics, Korea University, Seongbuk-gu, Seoul, Republic of Korea}

\author{Wieland Sch\"ollkopf}
\affiliation{Fritz-Haber-Institut
der Max-Planck-Gesellschaft, Faradayweg 4-6, 14195 Berlin, Germany}
\author{Bum Suk Zhao (조범석)}
\email{zhao@unist.ac.kr}
\affiliation{Department of Physics, Ulsan
National Institute of Science and Technology (UNIST), Ulsan 44919, Republic of Korea}
\affiliation{Department of Chemistry, Ulsan
National Institute of Science and Technology (UNIST), Ulsan 44919, Republic of Korea}
\date{\today}

\begin{abstract} 
Thermal energy atom scattering at a surface with grazing incidence conditions is an innovative method for investigating dispersive atom-surface interactions with potential application in quantum sensing interferometry. The complete establishment of this technique would require a detailed peak analysis, which has yet to be achieved. We examined peak-width fluctuations in atomic and molecular beams diffracted by a grating under grazing incidence conditions. Careful measurements and analyses of the diffraction patterns of He atoms and D$_2$ molecules from three square-wave gratings with different periods and radii of curvature enabled the identification of factors influencing the variations in the width as a function of the grazing incidence angle. The effects of macroscopic surface curvature, grating magnification, and beam emergence are substantial under these conditions but negligible for incidence angles close to the normal. Our results shed light on the phenomena occurring in grazing incidence thermal energy atom scattering.
\end{abstract}

\maketitle

X-ray, electron, neutron, and atom diffraction techniques are well-established methods for studying the crystal structures of materials and their changes over time. In diffraction experiments, the width of the diffraction peak, along with its intensity and position, is a critical parameter for sample analysis. For example, peak-width analysis has been employed to estimate crystallite or grain sizes and crystal strains in X-ray powder diffraction \cite{Scherrer1918, Langford1978, Balzar2004, Holzwarth2011} and in grazing incidence X-ray scattering \cite{Smilgies2009, Mahmood2020}. In thermal-energy atom scattering (TEAS), the broadening of peak widths provides insights into temperature-induced alterations in surface morphology \cite{Farias98} and the density of defects such as steps on a crystal surface \cite{Poelsema1982, Poelsema1989}.

The design of optical elements such as mirrors and gratings for X-ray and matter-wave optics also necessitates a comprehensive investigation of the widths of scattering peaks. X-rays have been focused efficiently using cylindrical concave mirrors by minimizing peak broadening effects \cite{Kirkpatrick48, Mimura2010, Yumoto2013}. Recently, this endeavor has also been extended to He atoms \cite{Schewe2009}. Furthermore, understanding wavelength-dependent peak width broadening is essential for atom monochromators \cite{Schief1996}. Thus, analyzing peak widths is crucial for developing new methodologies and technologies based on wave diffraction. 

Grazing incidence thermal energy atom scattering (GITEAS) at a surface offers a unique approach that can complement conventional TEAS, akin to the relationship between X-ray scattering and grazing incidence X-ray scattering. The lower effective energy and longer wavelength toward the surface for GITEAS make it more sensitive to weak interactions and less responsive to surface roughness. As a result, GITEAS has become valuable for studying the dispersive interaction of atoms with a surface \cite{DeKieviet03, Zhao2008, Zhao2010a}. Furthermore, microstructure-grating interferometry with GITEAS can be applied for quantum sensing in conjunction with monolithic atom interferometry using TEAS \cite{Fiedler2023}.

The versatile applications of GITEAS necessitate a precise peak-width analysis. However, under grazing incidence conditions, the peak widths are strongly influenced by an infinitesimal curvature (curvature radius of a few kilometers) and the diffraction direction near a surface, which results in unusual variations in peak widths. The presence of abnormally wide or narrow peaks further complicates the analysis. Furthermore, the traditional peak-width analysis scheme used in TEAS is insufficient for GITEAS, which highlights the need for a more sophisticated approach. 

In this letter, we report a comprehensive analysis of peak widths for GITEAS. By adjusting the grating and incident beam properties, we investigate various factors contributing to peak-width variations, such as the macroscopic surface curvature, grating magnification, incident beam divergence, and angular dispersion. He atoms (D$_2$ molecules) with mean de Broglie wavelengths $\lambda$ of 330 or 140 pm (140 pm) are diffracted at grazing incidence angles up to 30 mrad by three gratings of different periods and macroscopic curvatures. By comparing the measured peak widths with calculated widths, we identify the dominant factors influencing the variations in peak width. This resolves any potential ambiguities in the data analysis caused by extraordinary peak widths.

%
\begin{figure*}[t]
\includegraphics[scale=0.45]{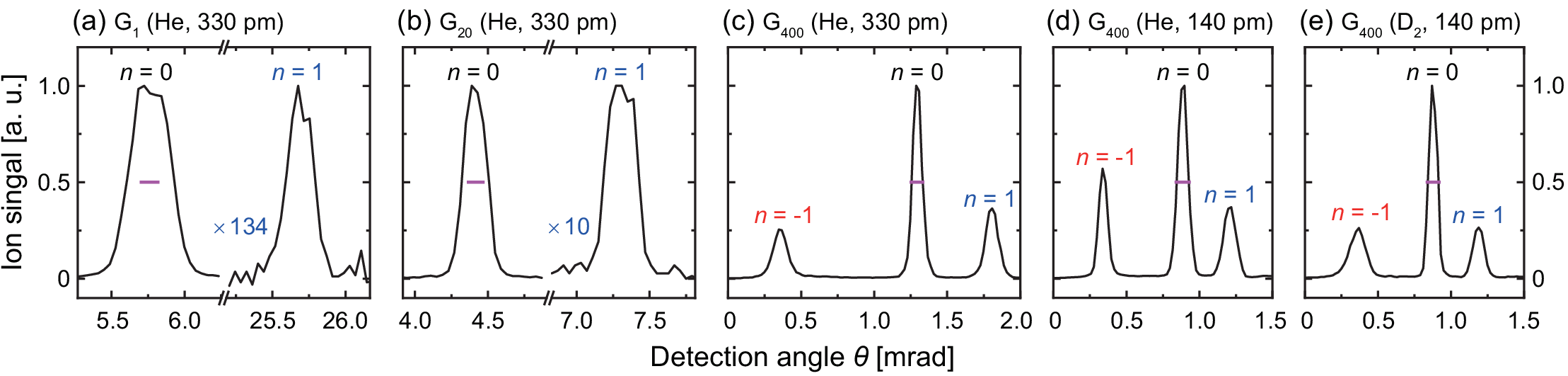}
\caption{Measured angular spectra of matter-wave beams of He (a-d) and D$_2$ (e) diffracted by grating G$_1$ (a), G$_{20}$ (b), and G$_{400}$ (c-e) at incidence angles $\theta_{\rm in}$ of (a) 5.77, (b) 4.41, (c) 1.29, (d) 0.89, and (e) 0.88 mrad. The horizontal bar near the specular peak indicates the corresponding width $w_{\rm in}$ of the beam with no grating installed. In (a) and (b), the 1st-order peak is magnified for clarity.}
\label{fig:spectra}
\end{figure*}

Our atom optics apparatus characterized by a tightly collimated incident beam and high-angular-resolution detector enables precise measurements of peak widths. Further details of the setup can be found in previous references \cite{Zhao2008, Zhao2019} and the Supplemental Material \cite{supp2}. Here, we focus on the aspects of the apparatus pertinent to the data analysis presented in this study. A continuous beam of He or D$_2$ is formed by supersonic expansion of the corresponding pure gas from a source cell. The source temperature ($T_0$) and pressure ($P_0$) influence the particle velocity distributions. We use three sets of source conditions, viz., gas species, $T_0$, and $P_0$: (He, 9.0 K, 0.5 bar), (He, 52 K, 26 bar), and (D$_2$, 52 K, 2 bar). For each set of conditions, we observe a velocity distribution, from which the mean velocity ($v$), full width at half maximum (FWHM) ($\Delta v$), and corresponding mean de Broglie wavelength $\lambda$ (see Fig.~SM2 \cite{supp2}) are determined. Accordingly, we obtain three corresponding sets of incident beam properties, including $v$, $\Delta v$, and $\lambda$: (304 m/s, 2.3 m/s, 330 pm), (733 m/s, 5.9 m/s, 140 pm), and (736 m/s, 79 m/s, 140 pm), respectively, which are used to explore the effects of $\lambda$ and $\Delta v$ separately. 

The beam is collimated using two slits (S1 and S2) positioned 100 cm apart as shown in Fig.~SM1 \cite{supp2}. The widths of these slits, $W_{\rm S1}$ and $W_{\rm S2}$, are 20 $\mu$m, except for one set of data where $W_{\rm S1} = 10$ $\mu$m and $W_{\rm S2} = 15$ $\mu$m. The incident and scattered beams are detected by precisely rotating a mass spectrometer with electron-bombardment ionization. The rotational axis of the detector is located 40 cm downstream from S2. A third slit (S3) with a width of $W_{\rm S3} = 25$ $\mu$m is positioned just before the detector. The distance between the rotational axis and S3, referred to as the grating--detector distance, is $L = 38$ cm. 

We employ three square-wave gratings with varying periods $d$ and strip widths $a$: G$_1$ with $d=1$ $\mu$m and $a=0.25$ $\mu$m; G$_{20}$ with $d=20$ $\mu$m and $a=10$ $\mu$m; and G$_{400}$ with $d=400$ $\mu$m and $a=200$ $\mu$m. Albeit nominally plane, the grating surfaces exhibit small circular curvatures. Under grazing incidence conditions the curvature along the direction perpendicular to the incident plane affects the peak width only negligibly \cite{Erko2008, Als-Nielsen2011}. Therefore, we consider the gratings as cylindrical with their curvature radii $R$. For concave and convex gratings, $R>0$ and $R<0$, respectively. The estimated $R$ values of G$_1$, G$_{20}$, and G$_{400}$ are 30, $-210$, and 1800 m, respectively. 


Figure \ref{fig:spectra} shows angular spectra for the three gratings measured at different experimental conditions for various incidence angles. The graphs represent the He$^+$ or D$_2^+$ signal as a function of the detection angle $\theta$. Here, $\theta_{\rm in}$ and $\theta$ are measured with respect to the grating surface. Numbers $n$ indicate the diffraction order assigned to each peak. The peak positions $\theta_n$ and FWHM values $w_n$ of the $n$th-order diffraction beams are determined by fitting each peak to a single Gaussian function. Similarly, we determine the FWHM $w_{\rm in}$ of the incident beam spectrum when the grating is removed from the beam path.

A peak-width analysis reveales unexpected irregular hierarchies of $w_n$ for the 5 cases of Fig \ref{fig:spectra}: (a) $w_0 > w_1 > w_{\rm in}$, (b) $w_1 > w_0 > w_{\rm in}$, (c) $w_{-1} > w_{\rm in} \approx w_1 > w_0 $, (d) $w_{\rm in} > w_1 > w_0 > w_{-1}$, and (e) $w_{-1}> w_{\rm in} > w_1 > w_0$. Generally, in TEAS, diffraction peak widths increase with $|n|$ owing to the angular dispersion \cite{Estermann30, Grisenti2000}, and an increase of the specular width $w_0$ with respect to $w_{\rm in}$ can be attributed to surface defects \cite{Poelsema1981}. Therefore, the unexpected hierarchies could lead to misinterpretation of underlying physics and errors in peak assignment.

\begin{figure*}[t]
\includegraphics[scale=0.45]{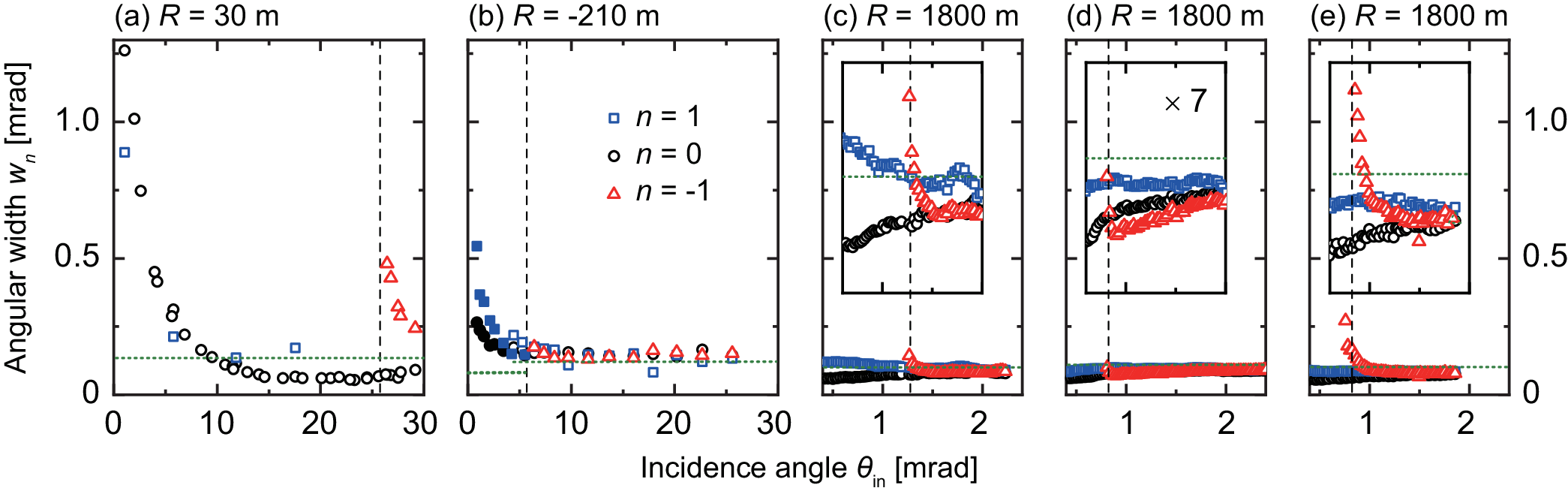}
\caption{Angular FWHM of the $n$th-order diffraction peak, $w_n$, as a function of the incidence angle $\theta_{\rm in}$. The experimental conditions are the same as those in Fig.~\ref{fig:spectra}. The vertical dashed line denotes the Rayleigh incidence angle of the $-1$st-order diffraction beam, referred to as $\theta_{{\rm R},-1}$, where the $-1$st-order diffraction beam emerges from the grating surface. The horizontal dotted line represents the width of the incident beam, $w_{\rm in}$. In (b), filled symbols within the range of $\theta_{\rm in}$ from 0.5 to 5.5 mrad illustrate data obtained with tight collimation, resulting in a narrower $w_{\rm in}$ of 0.081 mrad (thick horizontal line). The insets in (c)\textendash(e) show seven-fold magnifications of the corresponding data series.} \label{fig:width}
\end{figure*}

To study the peak-width variations systematically, we plot $w_n$ as a function of $\theta_{\rm in}$ for the five experimental conditions in Fig.~\ref{fig:width}. Each graph includes horizontal dotted lines indicating $w_{\rm in}$ and vertical dashed lines representing the Rayleigh incidence angle of $-1$st-order diffraction beam emergence ($\theta_{{\rm R},-1}$). When $\theta_{\rm in} =\theta_{{\rm R},-1}$, the $-1$st-order diffraction beam emerges from the grating and propagates parallel to its surface; in this case, $\theta_{-1}=0$ \cite{Zhao2010, Zhao2011a}.

The relationship between $w_n$ and $\theta_{\rm in}$ varies under different experimental conditions and for individual diffraction orders. Furthermore, the inconsistent hierarchies among $w_0$, $w_1$, and $w_{-1}$ change with $\theta_{\rm in}$. Several factors contribute to these variations: (i) the macroscopic curvature of the grating surface, (ii) grating magnification, (iii) diffraction beam spread resulting from the divergence of the incident beam, and (iv) angular dispersion due to the non-monochromatic nature of the beam. Among these factors, (i) pertains to a property of the grating, (ii) results from the diffraction principle, and (iii) and (iv) are determined by the incident beam properties. 

The macroscopic curvature of the grating surface accounts for the variation in $w_0$ shown in Fig.~\ref{fig:width}. The magnitude of $|R|$ directly influences the steepness of the decrease in $w_0$. Additionally, when $R>0$ ($R<0$), $w_0$ increases (decreases) asymptotically toward $w_{\rm in}$ with $\theta_{\rm in}$ as illustrated in Fig.~\ref{fig:width}(a) [Fig.~\ref{fig:width}(b)]. Furthermore, the steep increase in $w_{-1}$ with decreasing $\theta_{\rm in}$ in Fig.~\ref{fig:width}(a) and ~\ref{fig:width}(e) is attributed to angular dispersion. Finally, the hierarchical order of $w_1 > w_0 > w_{-1}$ shown in the inset of Fig.~\ref{fig:width}(d) results from the grating magnification.  

Factors (i)\textendash(iv) contribute to $w_n$ differently depending on $n$, $\lambda$, and $\theta_{\rm in}$, which can be formulated by the following approximate equation for a linear width $W_n = L w_n$: $W_n= \sqrt{[W_n^{(1)}]^2+[W_n^{(2)}]^2+[W_n^{(3)}]^2+[W^{(4)}_n]^2+[W^{(5)}]^2}$, where
\begin{equation}\label{eq:eq1}
   W^{(1)}_n=\left( \frac{1}{o}+\frac{1}{L}-\frac{1}{f_n}\right) L W_{\rm G}\frac{\sin \theta_n}{\sin \theta_{\rm in}}
   \, ,
\end{equation}
\begin{equation}\label{eq:eq2}
   W^{(2)}_n=\frac{2.355}{4}\frac{W_{\rm S1}}{o}L\frac{\sin \theta_{\rm in}}{\sin \theta _n}
   \, ,
\end{equation}
\begin{equation}\label{eq:eq3}
   W^{(3)}_n=0.884\frac{\lambda}{W_{\rm S2}}L\frac{\sin \theta_{\rm in}}{\sin \theta _n}
   \, ,
\end{equation}
\begin{equation}\label{eq:eq4}
   W^{(4)}_n=\left( \frac{|n|\lambda}{d}\right) \frac{\Delta v}{v}\frac{1}{\sin \theta_n}L
   \, ,
\end{equation}
and
\begin{equation}\label{eq:eq5}
   W^{(5)}=\frac{2.355}{4}W_{\rm S3}
   \, .
\end{equation}
In these equations, $o$ represents the object distance, $f_n$ denotes the focal length of the $n$th-order diffraction beam, and $W_{\rm G}$ denotes the width of the incident beam at the center of the grating. Because S1 constrains the effective source size of the matter-wave beam, we approximate the object as a Gaussian distribution with a standard deviation of W$_{\rm S1}$/4. Similarly, the boxcar-shaped function defined by S3 is approximated as a Gaussian function with a standard deviation of W$_{\rm S3}$/4.

The macroscopic curvature of the grating represented by $R$ is relevant to its focal length. Under grazing incidence conditions, the object distance $o$ and image distance $i_n$ of the $n$th-order beam satisfy the thin lens equation: $1/o+1/i_n = 1/f_n$. The term in the parentheses of Eq.~(\ref{eq:eq1}) then represents the focusing error, $\epsilon_n$, and the product $\epsilon_n L W_{\rm G}$ is the width of the (de-)focused incident beam at the detection plane \cite{Saleh2019}. The focal length $f_n$ is a function of $R$ and $\theta_{\rm in}$. Specifically, $f_0$ can be expressed as $f_0 = R\sin \theta_{\rm in}/2$ \cite{Kirkpatrick48,Schewe2009}; therefore, $i_n$ (or $\epsilon_n$) can be determined based on the values of $R$ and $\theta_{\rm in}$, as illustrated in Fig.~SM4 \cite{supp2}.

The grating magnification given by $M_n = \sin \theta_n / \sin \theta_{\rm in}$, also known as anamorphic magnification, represents the ratio of the width of a collimated diffracted beam to that of a collimated incident beam \cite{Palmer2020}. When considering a collimated beam ($o \rightarrow \infty$) incident on a flat grating ($f_n \rightarrow \infty$), $W^{(1)}_n$ characterizes the grating magnification. 

\begin{figure}[t]
\includegraphics[scale=0.4]{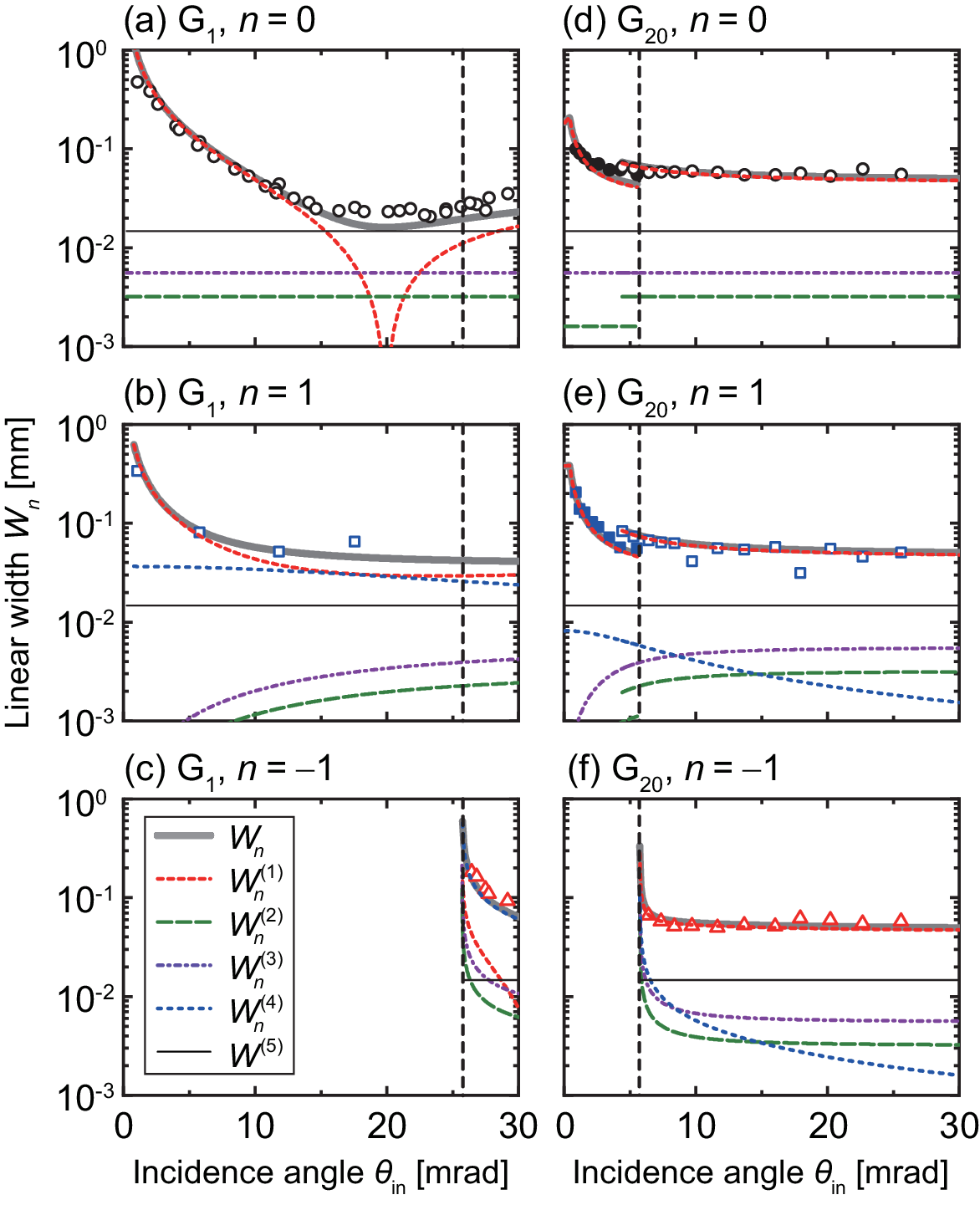}
\caption{Comparison between the measured (symbols) and calculated (solid curves) linear widths $W_n$ of the $n$th-order diffraction beam. The first- and second-column data refer to G$_1$ and G$_{20}$, respectively. The data are listed sequentially from the top for $n=0$, 1, and $-1$. $W_0$, $W_1$, and $W_{-1}$ are plotted separately with the calculated $W^{(1)}_n$, $W^{(2)}_n$, $W^{(3)}_n$, $W^{(4)}_n$, and $W^{(5)}$.} \label{fig:width_g1}
\end{figure}

$W^{(2)}_n$ describes the effect of the geometrical incidence beam divergence $\simeq W_{\rm S1}/o$ on the diffraction beam spread $\Delta \theta_n$. When $W_{\rm S1}=20$ $\mu$m, $W^{(2)}_n=3.2$ $\mu$m for the specular peak, which is negligible. In contrast, $W^{(2)}_n$ becomes significant when $\theta_{n<0} \rightarrow 0$ as it happens when $\theta_{\rm in}$ approaches the Rayleigh angle of beam emergence $\theta_{{\rm R},n}$.

Slit diffraction at S2 contributes additional incident-beam divergence $\Delta \theta_{\rm in, SD}$, which is responsible for $W^{(3)}_n$. In the Fraunhofer limit, $\Delta \theta_{\rm in,SD} = 0.844 \lambda / W_{\rm S2}$. Similar to $W^{(2)}_n$, this contribution becomes pronounced only for an emerging peak close to the Rayleigh condition.

$W^{(4)}_n$ accounts for diffraction peak broadening resulting from angular dispersion due to the beam's finite velocity spread $\Delta v$. This effect is absent in the specular peak. 

Finally, the finite size of the detector entrance slit, S3, also contributes to the observed diffraction peak widths by the term ($W^{(5)}$) of Eq.~(\ref{eq:eq5}). This contribution is a constant 15 $\mu$m for all experimental conditions. 

To assess the relative contributions of these five terms to $W_n$, we compare the measured $W_n$ (symbols) for G$_1$ and G$_{20}$ with the corresponding calculated values for $W_n$, $W^{(1)}_n$, $W^{(2)}_n$, $W^{(3)}_n$, $W^{(4)}_n$, and $W^{(5)}$ (lines) in Fig.~\ref{fig:width_g1}. The theoretical curves are determined considering the grating's macroscopic curvature radius $R$ as the sole adjustable parameter. The angular width $w_n$ presented in Figs.~\ref{fig:width}(a) and \ref{fig:width}(b) is converted to the linear width $W_n$ and presented in the first- and second-column graphs in Fig.~\ref{fig:width_g1}, respectively. The values for $W^{(1)}_n$ corresponding to G$_1$ (G$_{20}$) are calculated with $W_{\rm G} = 38.5$ $\mu$m ($W_{\rm G} = 34.2$ $\mu$m). The estimation of $W_{\rm G}$ employs the linear width of the incident beam, namely, $W_{\rm in}=51.2$ $\mu$m ($W_{\rm in}=46.0$ $\mu$m). 

For G$_{20}$ a second data set was obtained with tighter beam collimation ($W_{\rm S1} = 10$ $\mu$m, $W_{\rm S2} = 15$ $\mu$m) at small incidence angles $\theta_{\rm in} < 5.5$ mrad (filled symbols in Fig.~\ref{fig:width}(b)) resulting in an angular width as small as $w_{\rm in}$ = 0.081 mrad ($W_{\rm in} = 30.8$ $\mu$m and $W_{\rm G} = 21.2$ $\mu$m). Figures \ref{fig:width_g1}(d)\textendash\ref{fig:width_g1}(f) illustrate the corresponding calculations.

The breakdown of $W_n$ into its five constituent terms in Fig.~\ref{fig:width_g1} highlights the dominant factors in each case. The substantial reductions in $W_n$ for G$_1$ at varying incidence angles result from different factors for $n=0$, 1, and $-1$. As illustrated in the left-column graphs, at $\theta_{\rm in} < 10$ mrad, $\theta_{\rm in} < 5$ mrad, and $\theta_{\rm in} < 30$ mrad, $W^{(1)}_0$, $W^{(1)}_1$ (red dashed curves), and $W^{(4)}_{-1}$ (blue-dotted curve) predominantly influence their respective $W_n$ values. Notably, the grating magnification is unity for the specular beam, which makes $\epsilon_0$ the key determinant for $W^{(1)}_0$. Conversely, $\epsilon_1$ varies by less than 33$\%$ in the given range of incidence angles while the grating magnification term $M_1$ decreases tenfold [see Figs.~SM4(c1) and SM4(d1)] \cite{supp2}. Therefore, the principal contributors to the steep decline in $W_n$ for $n =0$, 1, and $-1$ are the macroscopic curvature, grating magnification, and angular dispersion, respectively.

Similar to G$_1$, the decreases in $W_0$ and $W_1$ for G$_{20}$ are primarily determined by the macroscopic curvature and grating magnification, respectively [see Figs.~\ref{fig:width_g1}(d), \ref{fig:width_g1}(e), SM4(c2), and SM4(d2)]. However, unlike G$_1$, the 20-fold larger $d$ of G$_{20}$ diminishes the effect of angular dispersion [see Eq.~(\ref{eq:eq4})]; as a result, the influence of curvature on the reduction of $W_{-1}$ becomes dominant. 

Even though G$_{400}$ is nearly flat with $R=1800$ m, given the extreme grazing-incidence conditions the curvature still affects the peak width variations. As can be seen in Figs.~\ref{fig:width}(c)\textendash\ref{fig:width}(e) all three peak widths $w_1$, $w_0$, and $w_{-1}$ can be narrower than the incidence beam width $w_{\rm in}$ resulting from beam focusing due to the concave curvature of the grating (refer to the first and second rows in Fig.~SM5) \cite{supp2}. 

In addition, the width hierarchy visible in Fig.~\ref{fig:width}(d) ($w_1 > w_0 > w_{-1}$) presents a clear example for peak widths being dominated by grating magnification; the closer a beam propagates to the surface, the smaller its width. This trend is less clear in Fig.~\ref{fig:width}(c) and (e) where $w_{-1}$ is not consistently the smallest width at the given incidence angles. Specifically, for incidence angles slightly larger than the $-1$-order Raleigh angle $\theta_{{\rm R},-1}$, $w_{-1}$ exceeds both $w_0$ and $w_1$.

As shown in Figs.~SM5(c), SM5(f), and SM5(i), the larger contributions of the terms $W^{(2)}_{-1}$, $W^{(3)}_{-1}$, and $W^{(4)}_{-1}$ compared with $W^{(1)}_{-1}$ lead to broadening of $W_{-1}$. The small $\theta_{-1}$ close to Rayleigh conditions boosts these three terms. Interestingly, as shown in Fig.~SM5, $W^{(2)}_{-1}$ and $W^{(3)}_{-1}$ were the dominant factors for the He atom beams with two different $\lambda$, whereas $W^{(4)}_{-1}$ was the crucial factor for the D$_2$ molecular beam. This behavior can be attributed to the 13-times larger velocity spread $\Delta v$ of the D$_2$ beam compared with that of the He beam at identical velocity $v$.

In conclusion, our combined experimental and theoretical investigations of diffraction peak widths in GITEAS reveal the primary factors that induce variations in peak widths. Notably, the primary factor governing the width of diffraction beams varies depending on diffraction order, incidence angle, and grating period. Our study reveals the effects of macroscopic curvature, emerging beams, and grating magnification, which have been overlooked in other scattering techniques, such as TEAS. Thus, our findings address potential ambiguities in interpreting diffraction data as those presented in Fig.~\ref{fig:spectra}. 

The comprehensive peak-width analysis conducted in this study lays the groundwork for extending the applicability of GITEAS to investigate the unique characteristics of dispersive interactions between atoms and thin-layer surfaces such as graphene sheets or few layers hexagonal boron nitride, known for their flexibility. While recent theoretical investigations \cite{Woods2016} have delved into these interactions, limited experimental studies are available. Additionally, this analysis can guide the design of atom optical components. Although both $w_{1}$ at $\theta_{\rm in} = 3$ mrad and $w_{-1}$ near $\theta_{{\rm R},-1}$ in Fig.~\ref{fig:width}(a) are sufficiently broad for monochromator applications, only the $-1$st-order diffraction beam is suitable for this purpose. This is because wavelength-dependent angular dispersion and wavelength-independent grating magnification primarily influence $w_{-1}$ and $w_1$, respectively. Furthermore, peak-width analysis will become critical in atom interferometry using GITEAS.

This study was supported by NRF (National Research Foundation of Korea) grants funded by the Korean Government (NRF-2020R1A2C3003701 and NRF-2022M3C1C8094518).

L.~Y.~K.~and D.~W.~K.~contributed equally to this work.

\pagebreak

\clearpage

\noindent \Large\textbf{Supplemental Material}
\bigskip

\renewcommand{\thefigure}{SM\arabic{figure}}
\renewcommand{\thetable}{SM\arabic{table}}
\renewcommand{\theequation}{SM\arabic{equation}}
\setcounter{figure}{0}
\setcounter{equation}{0}

\normalsize
\noindent\textbf{Atomic and molecular beam apparatus}
\newline
\begin{figure}[b]
\includegraphics[scale=0.4]{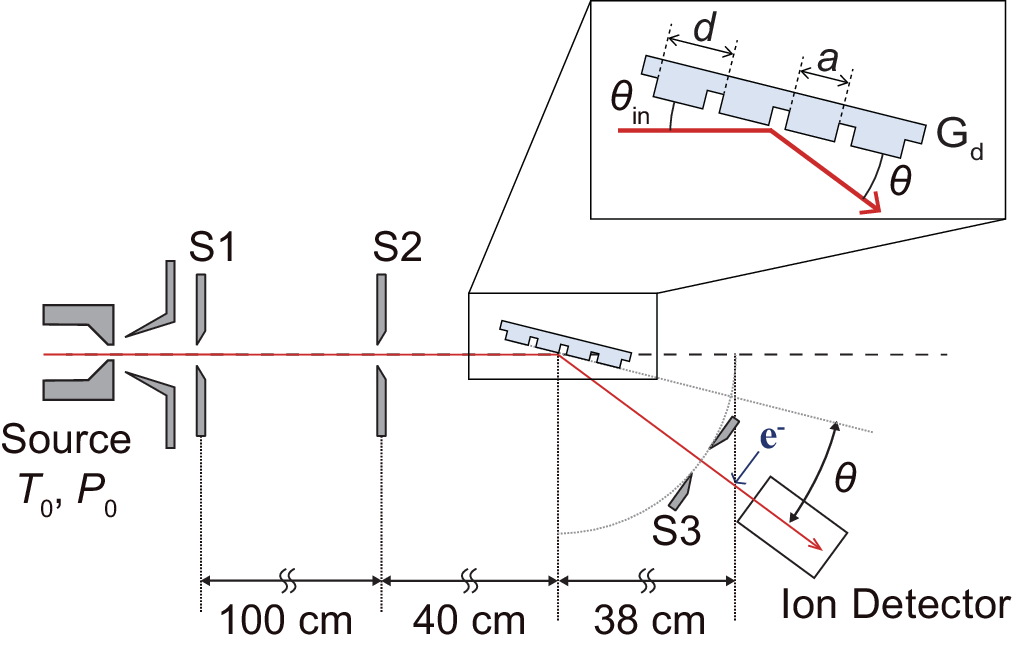}
\caption{Schematic of the experimental setup. The distances between the components are not drawn to scale. The incidence and detection angles, $\theta_{\rm in}$ and $\theta$, are measured with respect to the grating surface. The inset illustrates a grating with $d$ and $a$ representing its period and strip width, respectively.}
\label{fig:setup}
\end{figure}

A continuous beam was generated by supersonically expanding either He atoms or D$_2$  molecules through a 5-$\mu$m-diameter nozzle. The temperature $T_0$ and pressure $P_0$ of the source were set to 9.0 K and 0.5 bar, respectively, for He atoms of $\lambda=330$ pm and set to $T_0 = 52$ K and $P_0= 26$ bar, respectively, for He atoms of $\lambda=140$ pm. To compare the diffractions of the He atomic and D$_2$ molecular beams of the same $\lambda$ values, we set the conditions for D$_2$ molecules to $P_0=2$ bar and $T_0 = 52$ K. The $\lambda$ values were determined from the velocity distributions of the particles. 

The beam first passed through a skimmer with a diameter of 500 $\mu$m and was subsequently collimated by two slits (S1 and S2) placed 100 cm apart. The widths of these slits, $W_{\rm S1}$ and $W_{\rm S2}$, were the same ($W_{\rm S1} = W_{\rm S2} = 20$ $\mu$m) for most measurements while for one set of data shown in this paper it was $W_{\rm S1} = 10$ $\mu$m and $W_{\rm S2} = 15$ $\mu$m. The incident and scattered beams were detected by precisely rotating a mass spectrometer with electron-bombardment ionization. The rotational axis of the detector was located 40 cm downstream from S2. Just before the ion detector, a third slit (S3) with a width of $W_{\rm S3} = 25$ $\mu$m was positioned to enhance the angular resolution. The distance between the rotational axis and S3, referred to as the grating--detector distance (or detector radius), was $L = 38$ cm. This distance, combined with the narrow aperture of S3, yielded a full width at half maximum (FWHM) of approximately 0.12 mrad for the collimated incident beam. The grating was carefully positioned and oriented such that the pivot axis of the detector aligned with the center of the grating plane and was maximally parallel to the grating grooves. Therefore, the grating grooves were perpendicular to the plane of incidence defined by the incident wave vector and grating normal, resulting in in-plane diffraction. This geometry is illustrated in Fig.~1 in the main text, where both the incidence angle $\theta_{\rm in}$ and detection angle $\theta$ are measured in relation to the grating plane.

\begin{figure}[t]
\includegraphics[scale=0.8]{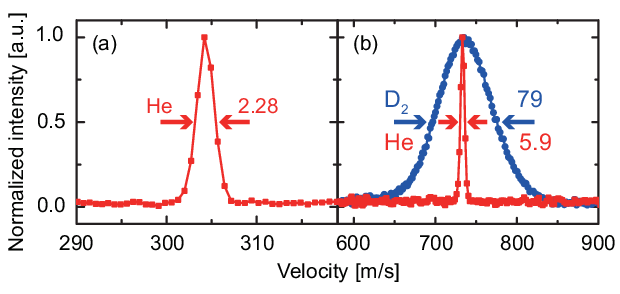}
\caption{Measured velocity distributions of He (red squares) and D$_2$ (blue circles) at (a) $T_0=$ 9 and (b) 52 K. The intensity is scaled to the peak values. Under conditions of $P_0 = 0.5$ bar and 26 bar for $T_0=$ 9 K and 52 K, respectively, the FWHM values  for the He atom beams are 2.28 and 5.9 m/s. In contrast, the velocity width of the D$_2$ molecular beam is 79 m/s at $T_0=$ 52 K. The mean velocities for these distributions are approximately 304 and 730 m/s for $T_0=$ 9 and 52 K, respectively.}
\label{fig:TOF}
\end{figure}

\bigskip
\noindent\textbf{Velocity distributions of atomic and molecular beams}
\newline
We obtained the velocity distributions of the He atomic and D$_2$ molecular beams using time-of-flight measurements conducted with a mechanical pseudo-random beam chopper \cite{ich:epjd04}. Figures \ref{fig:TOF}(a) and \ref{fig:TOF}(b) present the results obtained at temperatures of $T_0 = 9.0$ K and 52 K, respectively. We employed Gaussian function fitting to determine both the mean velocity $v$ and FWHM $\Delta v$. The ratios of $\Delta v/v$ for the He atom beams are 0.0075 and 0.0081 at the two different $T_0$ values, which are similar. In contrast, the ratio $\Delta v/v$ for the D$_2$ molecular beam is 0.11.

\bigskip
\noindent\textbf{Fabrication and characterization of the diffraction gratings}
\newline

\begin{table}[b]
\caption{Grating properties relevant to the peak-width variation}\label{tab:grating}
\begin{tabular}{lcrcr}
\hline
              && period && curvature radius  \\            
\hline
G$_1$  && 1 $\mu$m  && 30 m  \\
G$_{20}$ && 20 $\mu$m && $-210$ m \\
G$_{400}$&&400 $\mu$m && 1800 m \\
\hline
\end{tabular}
\label{table:grating}
\end{table}

We employed three square-wave gratings with varying periods $d$ and strip widths $a$: G$_1$ with $d=1$ $\mu$m and $a=0.25$ $\mu$m, G$_{20}$ with $d=20$ $\mu$m and $a=10$ $\mu$m, and G$_{400}$ with $d=400$ $\mu$m and $a=200$ $\mu$m. G$_1$ and G$_{20}$ are microstructured arrays, each measuring 56 mm in length, consisting of 110-nm-thick chromium stripes that are 5 mm in length and are deposited on a 2-mm-thick quartz surface. In contrast, G$_{400}$ is an array featuring parallel photoresist strips with a thickness of 1 $\mu$m, width of 200 $\mu$m, and length of 4 mm. These strips are located on a commercial gold mirror (Thorlabs PFSQ20-03-M03), which is 6 mm thick and has a surface area of $50.8 \times 50.8$ mm$^2$. Only the stripes were exposed to the incident atomic beam for all the angles of incidence examined in this study. Table \ref{table:grating} provides a summary of the properties of these gratings that are pertinent to variations in peak width, including their period $d$ and radius of curvature $R$.

\begin{figure}[t]
\includegraphics[scale=0.5]{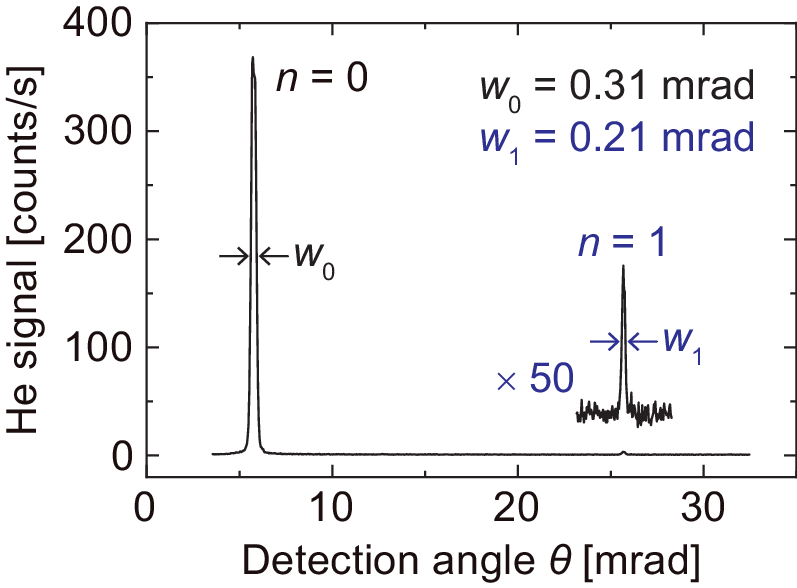}
\caption{Angular spectrum of the He atom beam diffracted from G$_1$ at $\theta_{\rm in}= 5.77$ mrad (the insets show parts of the spectrum on an enlarged scale).}
\label{fig:spectrum}
\end{figure}

\bigskip
\noindent\textbf{Example angular spectrum}
\newline
Figure \ref{fig:spectrum} shows the full angular spectrum of the He atom beam diffracted from G$_1$ at $\theta_{\rm in}= 5.77$ mrad [Fig.~1(a)]. The specular peak position $\theta_0$ equals $\theta_{\rm in}= 5.77$ mrad. The widths of specular and 1st-order-diffraction peaks ($w_0$ and $w_1$) are 0.31 mrad and 0.21 mrad, respectively. 

\bigskip
\noindent\textbf{Experimental conditions}
\newline
Table \ref{table:conditions} lists the five experimental conditions E1–E5 at which the five data sets presented in this study were obtained.

\begin{table}[h]
  \caption{Experimental conditions employed in this experiment }
    \begin{tabular}{ccccccccccccc}
    \hline
    &     && Particle && Grating        && $\lambda_{\rm dB}$ && $\textit{T}_0$ && $\textit{P}_0$ & \\
    \hline
    & E1 &&  He    && G$_1$   && 330 pm          && 9.0 K        && 0.5 bar & \\
    & E2 &&  He    && G$_{20}$  && 330 pm          && 9.0 K        && 0.5 bar & \\
    & E3 &&  He    && G$_{400}$ && 330 pm          && 9.0 K        && 0.5 bar & \\
    & E4 &&  He    && G$_{400}$ && 140 pm          && 52 K         && 26 bar  & \\
    & E5 && D$_2$  && G$_{400}$ && 140 pm          && 52 K         && 2.0 bar & \\
    \hline
    \end{tabular}
\label{table:conditions}    
\end{table}

\begin{figure*}[t]
\includegraphics[scale=0.33]{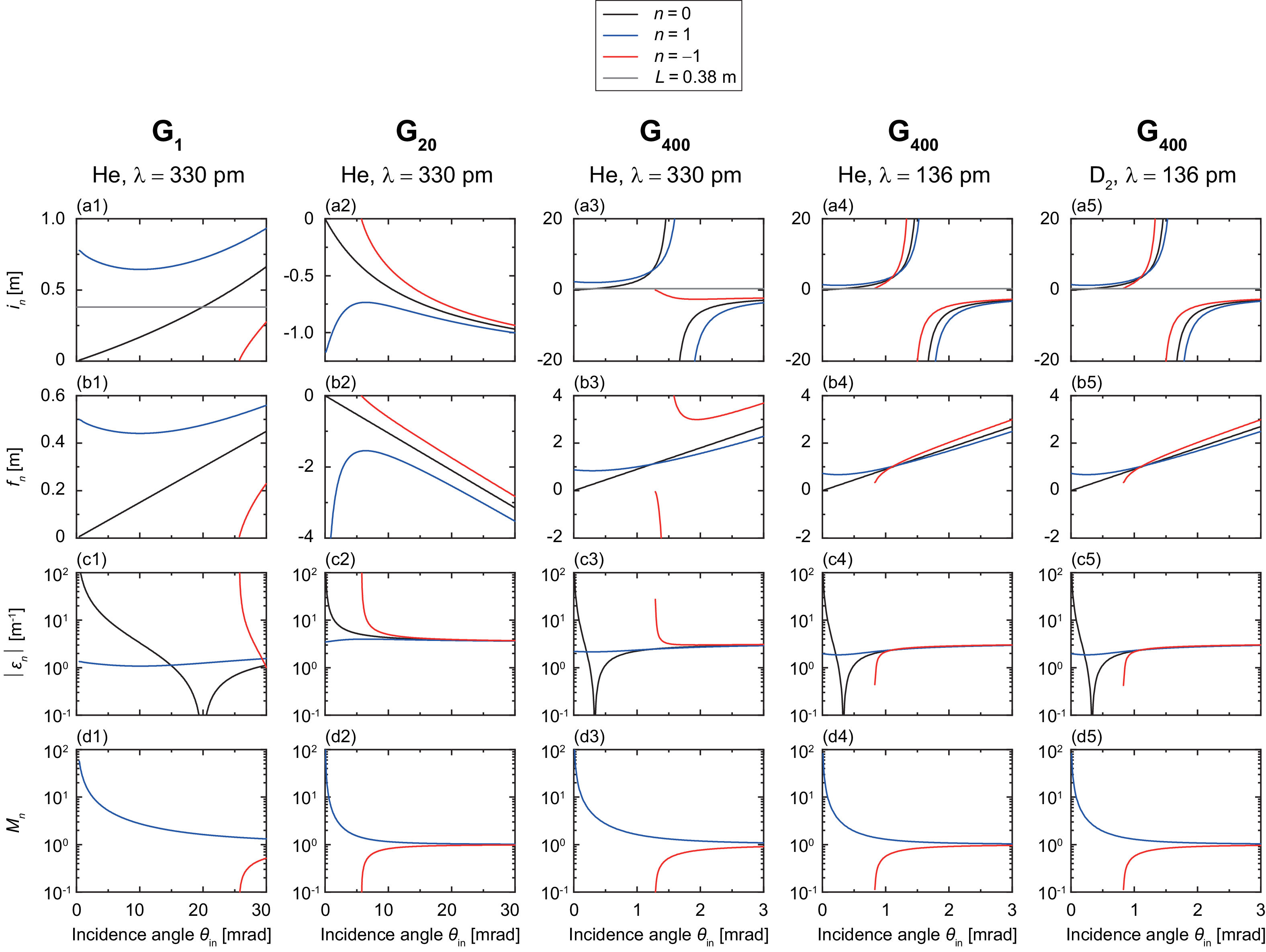}
\caption{(a1\textendash a5) Image distance $i_n$, (b1\textendash b5) focal length $f_n$, (c1\textendash c5) focusing error $\epsilon_n$, and grating magnification $M_n$ of the $n$th-order diffraction beam. The five columns correspond to the experimental conditions outlined in Table \ref{table:conditions}, respectively. Here, G$_1$, G$_{20}$, and G$_{400}$ are assumed to be cylindrical mirrors with curvature radii of $R=30$, $-210$, and 1800 m, respectively. These graphs are plotted as functions of the incidence angle $\theta_{\rm in}$. The horizontal dotted line in “(a1\textendash a5)” indicates the grating--detector distance $L = 0.38$ m.} 
\label{fig:b_image}
\end{figure*}

\begin{figure*}[t]
\includegraphics[scale=0.5]{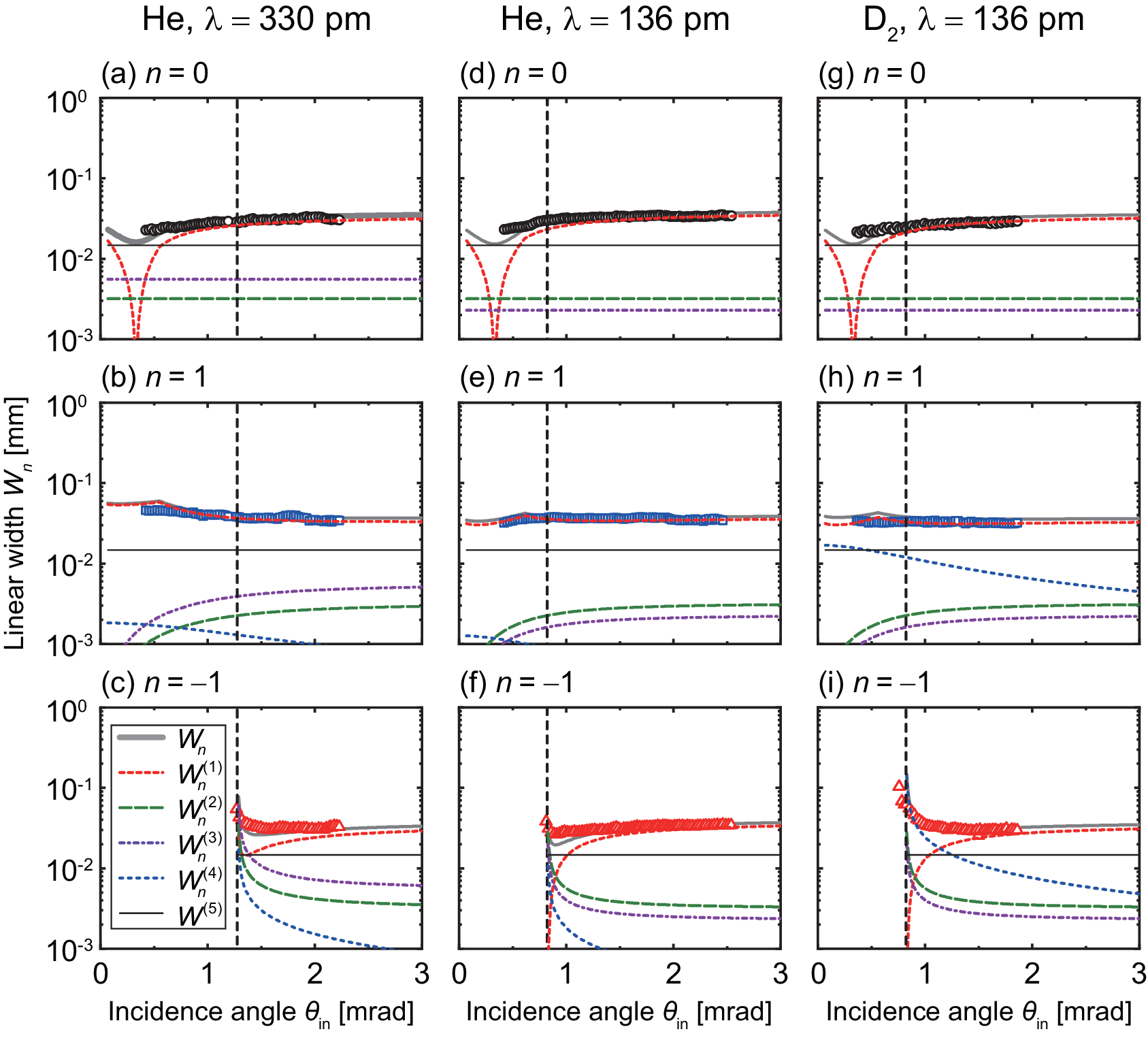}
\caption{Comparison between the measured (symbols) and calculated (solid curves) linear widths $W_n$ of the $n$th-order diffraction beam from G$_{400}$. The graphs in the three columns correspond to the experimental conditions E3, E4, and E5 as listed in Table \ref{table:conditions}, respectively. The data are presented sequentially from the top for $n=0$, 1, and $-1$. Specifically, $W_0$, $W_1$, and $W_{-1}$ are individually plotted alongside the calculated values $W^{(1)}_n$, $W^{(2)}_n$, $W^{(3)}_n$, $W^{(4)}_n$, and $W^{(5)}$.}
\label{fig:Width_G400}
\end{figure*}

\bigskip
\noindent\textbf{Calculation of parameters required for peak-width analysis}
\newline
For a cylindrically concave mirror with a curvature radius $R$, the object distance $o$ and image distance $i_n$ of the $n$th-order beam satisfy the following equation under grazing incidence conditions:
\begin{equation}\label{eq:eqS1}
   \frac{\theta_{\rm in}^+ - \alpha/2}{o-R\alpha} + \frac{\theta_n^+ + \alpha/2}{i_n+R\alpha} = \frac{2}{R}+\frac{(\theta_{\rm in}^{-}-\theta_n^{-})+(\theta_{\rm in}^{+}-\theta_n^{+})}{W_{\rm G}/\theta_{\rm in}}
   \, .
\end{equation}
Here, $\theta_{\rm in}^{+}$ and $\theta_{\rm in}^{-}$ represent the outermost values of the incidence angles that result in $\theta_n^{+}$ and $\theta_n^{-}$, respectively. The incidence beam spreads over a distance of $W_{\rm G}/\sin \theta_{\rm in}$ on the surface. This chord subtends an angle of $2\alpha$ at the center of curvature, making $\alpha$ approximately equal to $W_{\rm G}/(2R\sin \theta_{\rm in})$. Consequently, $i_n$ varies as a function of $\theta_{\rm in}$, as depicted in the first row of Fig.~\ref{fig:b_image} for the five experimental conditions. For our calculations, we assume $R=30$ m. When $i_n=L$, a diffraction beam is focused on the detection plane. Eq.~\ref{eq:eqS1} transforms into the thin-lens equation, $1/o+1/i_0=1/f_0$, for a specular beam of $n=0$ with $f_0 = R \sin \theta_{\rm in} /2 \simeq R \theta_{\rm in} /2 $ under grazing incidence conditions. Generally, $1/o+1/i_n=1/f_n$, with which we obtain $f_n$ using $i_n$ [See Figs.~\ref{fig:b_image}(b1\textendash b5)]. Then, $\epsilon_n=1/L-i_n$. The third row of Fig.~\ref{fig:b_image} shows $|\epsilon_n|$ for the five experimental conditions. To elucidate the contributions of the focusing error $\epsilon_n$ and the grating magnification $M_n$ to $W_n^{(1)}$, we plot $M_n$ in the last row of Fig.~\ref{fig:b_image}.

\bigskip
\noindent\textbf{Peak-width analysis for G$_{400}$}
\newline
Similar to Fig.~3 in the main text, Fig.~\ref{fig:Width_G400} compares the measured $W_n$ (symbols) for G$_{400}$ with the corresponding calculated values: $W_n$, $W^{(1)}_n$, $W^{(2)}_n$, $W^{(3)}_n$, $W^{(4)}_n$, and $W^{(5)}$ (solid lines). The theoretical curves are derived using $R=1800$ m. In most cases, the primary factor is $W^{(1)}_n$ (red dashed line). Exceptions occur for $W_{-1}$ at the Rayleigh conditions of $\theta_{\rm in} = \theta_{\rm R,-1}$. At these conditions, $\theta_{-1}$ approaches 0, leading to large values for $W^{(2)}_{-1}$, $W^{(3)}_{-1}$, and $W^{(4)}_{-1}$.

\end{document}